\documentclass{iaesarticle}
\usepackage[table,xcdraw]{xcolor}

\usepackage{amsmath, amsfonts, amssymb, float, fancyhdr}
\usepackage[figuresright]{rotating}
\usepackage{authblk, graphicx, indentfirst, lastpage, lipsum}
\usepackage{pifont}

\makeatletter
\renewcommand{\@biblabel}[1]{[#1]\hfill}
\renewcommand\AB@authnote[1]{\textsuperscript{\normalfont\bfseries#1}}
\makeatother
\usepackage[font=normalsize]{subfig, caption}
\usepackage{epstopdf}
\usepackage[left=3cm, right=2.5cm, top=2.5cm, bottom=2.5cm, includehead, includefoot]{geometry}
\usepackage[justification=centering]{caption}
\usepackage{titlesec}
\usepackage{xcolor}
\titleformat{\section}
{\normalfont\normalsize\bfseries\uppercase}{\thesection}{1.7em}{}
\titleformat{\subsection}
{\normalfont\normalsize\bfseries}{\thesubsection}{.95em}{}
\titleformat{\subsubsection}
{\bfseries}{\thesubsubsection}{.2em}{}
\titlespacing*{\section}{0cm}{0.7cm}{0cm}
\usepackage{enumitem}
\usepackage[numbers,sort&compress]{natbib}
\usepackage{ragged2e}
\usepackage{hyperref,graphicx}
\usepackage{multirow}

\CopyrightLine[]{}{\textit{This is an open access article under the \textcolor{blue}{\underline{CC BY-SA}} license.}
\vspace{.5em}}

\author[]{\bfseries Aminul Huq}
\author[]{\bfseries Md Tanzim Reza}
\author[]{\bfseries Shahriar Hossain}
\author[]{\bfseries Shakib Mahmud Dipto}
\affil[]{Department of Computer Science and Engineering, BRAC University, Dhaka, Bangladesh}

\title{AnoMalNet: outlier detection based malaria cell image classification method leveraging deep autoencoder}
\shorttitle{AnoMalNet: outlier detection based malaria cell image classification method leveraging ... (Aminul Huq)}

\begin{document}
\setcounter{page}{171}

\setlength{\parindent}{1.27cm}

\pagestyle{fancy}
\fancyhfoffset{0cm}

\journalname{International Journal of Reconfigurable and Embedded Systems (IJRES)}
\journalshortname{Int J Reconfigurable \& Embedded Syst}
\journalhomepage{http://ijres.iaescore.com}
\vol{13}
\no{1}
\months{March}
\years{2024}
\issn{2089-4864}
\DOI{10.11591}
\pagefirst{171}
\pagelast{178}

\maketitle

\hrule
\vspace{.1em}
\hrule
\vspace{.5em}
\noindent
\parbox[t][][s]{0.315\textwidth}{%
\textbf{Article Info}
\vspace{.5em}
\hrule
\vspace{.5em}
\begin{history}
\vspace{.5em}

Received Oct 10, 2022

Revised Jul 17, 2023

Accepted Aug 20, 2023

\vspace{.7em}
\end{history}
\vspace{.5em}
\hrule
\vspace{.5em}
\begin{keyword} 
\vspace{.5em}
Anomaly detection \sep 
Autoencoder \sep 
Class imbalance \sep 
Classification \sep
Malaria cell image \sep 
\vspace{.5em}
\end{keyword}
\vspace{\fill}
}
\parbox{0.020\textwidth}{\hspace{0.5em}}
\parbox[t][][s]{0.65\textwidth}{%
\begin{abstract}
\vspace{.3em}
Class imbalance is a pervasive issue in the field of disease classification from medical images. It is necessary to balance out the class distribution while training a model. However, in the case of rare medical diseases, images from affected patients are much harder to come by compared to images from non-affected patients, resulting in unwanted class imbalance. Various processes of tackling class imbalance issues have been explored so far, each having its fair share of drawbacks. In this research, we propose an outlier detection based image classification technique which can handle even the most extreme case of class imbalance. We have utilized a dataset of malaria parasitized and uninfected cells. An autoencoder model titled AnoMalNet is trained with only the uninfected cell images at the beginning and then used to classify both the affected and non-affected cell images by thresholding a loss value. We have achieved an accuracy, precision, recall, and F1 score of 98.49\%, 97.07\%, 100\%, and 98.52\% respectively, performing better than large deep learning models and other published works. As our proposed approach can provide competitive results without needing the disease-positive samples during training, it should prove to be useful in binary disease classification on imbalanced datasets.
\end{abstract}
}
\parbox[l]{\textwidth}{%
\rule{0.275\textwidth}{0.5pt} \hspace{0.5cm} \hrulefill
\\
\emph{\textbf{Corresponding Author:}}
\vspace{.5em}\\
Aminul Huq\\
Department of Computer Science and Engineering, BRAC University\\
66 Mohakhali, Dhaka, Bangladesh\\
Email: aminul.huq@bracu.ac.bd
}
\vspace{.5em}
\hrule
\vspace{.1em}
\hrule


\section{Introduction}

Malaria is a menacing disease that has affected large hordes of people in the past and continues to do the same at present. The statistics speak for themselves, 2020 saw a record-breaking estimation of 241 million malaria cases worldwide \cite{b1}. Needless to say, it is imperative to work on the remedies for such a deadly disease in order to mitigate the damages inflicted by them.

Microscopic thick and thin blood smear examinations are the most reliable and routinely used method for disease diagnosis. Thin blood smears help identify the species of the parasite causing the infection, whereas thick blood smears help detect the presence of parasites. However, the efficiency of this manual analysis method heavily depends on the medical personnel carrying out the tasks, also each diagnosis takes a huge deal of time. In the era of automation, where researchers are continuously working on fast and efficient ways of treating malaria, deep learning has been quite popular in terms of the detection and analysis of malaria as discussed in the literature review later on. Along with malaria classification, deep learning has held its grip in several research fields such as medical image analysis, natural language processing, and audio processing \cite{b2}-\cite{b10}

\newpage
Though deep learning techniques have proved to be handy in the recent past, it is observed that most of these models are heavy and use up a lot of computational power, this is also an issue when we try to deploy these models to mobile or edge devices. Additionally, these methods do not exhibit much potential when it comes to solving class imbalance issues. To rectify such issues this research work introduces AnoMalNet, an autoencoder-based method for the investigation of malaria in cell tissues that are lighter and capable to solve class imbalance problems in datasets. In this research the following contributions are made:
\begin{itemize}[topsep=0.3ex, itemsep=0.3ex, parsep=0.2ex, leftmargin=2ex]
  \item[-] An anomaly detection-based approach which is built upon autoencoders for the investigation of malaria in cell tissues that deals with class imbalance issues has been introduced.
  \item[-] The proposed model outperforms state-of-the-art models like VGG16, Resnet50, MobileNetV2, and LeNet.
  \item[-] Comparative analysis with other published methods has been provided.
\end{itemize}

\section{Literature Review}
There have been quite a few research in the field of malaria cell image classification. In the initial section of the review, we are going to discuss some of those. Later on we will have a short discussion on various techniques that have been used for handling class imbalance issue.

Raihan and Nahid \cite{b11} used a bioorthogonal wavelet to reduce the image size to 72$\times$72 resolution and extract the features . Images were passed through a custom convolutional neural network (CNN) with three convolutional layers and three fully connected (FC) layers. This CNN was used to extract the features from the first FC layer. In this way, 768 features were found initially. The whale optimization algorithm (WOA) was used to select the optimal subset of features. Samples of these features were passed through the XGBoost algorithm. For set 1 with 768 features, XGBoost achieved 94.92\%, 94.34\%, 95.57\%, and 94.95\% and for set 2 with 365 features, the model achieved 94.78\%, 94.39\%, 95.21\%, and 94.80\% for accuracy, precision, recall, and F1 score respectively in the validation set. XGBoost model construction time was half for the second set due to a number of features being reduced. Shapley additive explanations (SHAP) was used as a model explainability tool to assess the importance of the features \cite{b12}.

In the research article of Narayannan \emph{et al}. \cite{b13}, reshaped the images into 50$\times$50 resolution and the color consistency technique was applied to maintain the same illumination condition for all the images. A fast CNN model with 6 convolution layers and 2 FC layers was deployed. Additionally, AlexNet, ResNet, VGG-16, and DenseNet with transfer learning from imagenet were deployed. Furthermore, the bag-of-features model using SVM was used. Among all these implemented models, DenseNet got the highest accuracy of 96.6\%. Meanwhile, Reddy and Juliet \cite{b14} used a pre-trained ResNet with a sigmoid-enabled FC layer as the last layer. Apart from the last few layers, all the other layers were frozen during training. They achieved accuracies of 95.91\% and 95.4\% for training and validation respectively. The authors reported the existence of a test set in the experiment but no test result was reflected on it. 

Rajaraman \emph{et al}. \cite{b15} introduced a customized model with three convolution layers and two FC layers in their research work. AlexNet, Xception, ResNet, and DenseNet121 were used to extract features whereas gridsearch was used for hyperparameter optimization. For each individual CNNs, their default input resolution was used and for the pre-defined architectures, they tried extracting features from different layers and determined the most optimal layer to extract features from in order to improve accuracy. With extracted features from the most optimal layer, they got the highest accuracy of 95.9\% from VGG16 and from ResNet50 among all the tested models. In another research, thick blood smear images were collected \cite{b16}. Among those, 7,245 bounding box instances of plasmodium were annotated in 1,182 images. Images were divided into small patches and passed through a CNN to learn whether the patch contains any object of interest or not. The CNN was run on a 50/50 train-test split with which, they achieved an area under the receiver operating characteristic curve(ROC AUC) of 1.00. Authors claimed that their process is quite efficient since it can directly learn from pixel data. Furthermore, Bibin \emph{et al}. \cite{b17} introduced a trained model based on a deep belief network (DBN) to classify 4,100 peripheral blood smear images into two classes. By stacking limited Boltzmann machines and utilizing the contrastive divergence approach, the DBN is pre-trained. They took features from the photos and initialized the DBN's visible variables in order to train it. This paper's feature vector combines the attributes of color and texture. With an F-score of 89.66\%, a sensitivity of 97.60\%, and a specificity of 95.92\%, the proposed method has surpassed existing state-of-the-art methods significantly.

Lipsa and Dash \cite{b18} used the optimum number and size of convolution layers and spooling layers coupled with CNN. An Adam optimizer was employed to train and validate the model where a case study of the malaria diagnosis dataset was observed. Images were fed into the CNN keeping their size or color unchanged and assessments of their performance were made. An architectural comparison was performed between the proposed CNN model and some popular CNN architectures with the proposed model having a smaller number of hyperparameters. This comparison demonstrates that the mechanism of this model demands much fewer evaluation parameters, making the suggested approach a time-effective and computationally precise model in terms of predicate accuracy. Nugroho and Nurfauzi \cite{b19} used green, green, blue (GGB) color normalization as a preprocessing step in the detection of malaria. The findings demonstrate that their method has greater sensitivity and consistently comparable precision in a number of intersections over union (IoU) thresholds for malaria identification. Finally, Tan \emph{et al}. \cite{b20} employed an automated segmentation of one of the types, plasmodium falciparum out of 5 common types of malaria on a thin blood smear. It was experimented with using their proposed residual attention U-net. When the trained system was applied to verified test data, the results indicated an accuracy of 0.9687 and a precision of 0.9691.

Of all the research work that has been discussed so far, pretty much all of them used regular supervised learning-based methods on a balanced dataset. As discussed earlier, class imbalance in medical image datasets is not a rare incident and various approaches are taken to handle class imbalance. One prominent way is to generate synthetic data for the minority distribution class through a generative adversarial network (GAN) and balance out the class distribution. However, Mariani \emph{et al}.~\cite{b21} mentioned, GAN itself takes lots of images to generate synthetic data. Therefore, when the synthetic data generation is for a class that is sparse in distribution, it is not realistic to generate good-quality synthetic data since it is not possible to provide GAN with a sufficient amount of training images in the first place. Therefore, although GAN can hypothetically solve class imbalance issues, it is really hard to train a GAN in practice. Additionally, another common way of handling class imbalance is data augmentation. Well-known augmentation techniques include geometric transformation, noise injection, color space transformation, image mixing, applying kernel filter, cropping, random erasing, and so on~\cite{b22}. Some of these techniques, for example, image mixing and applying kernel filter may completely distort an image and change the underlying feature space. This feature transformation is generally unwanted in the case of medical images, as images of different modalities come with a very specific set of features. Additionally, other augmentation tactics, such as geometric transformation, cropping, and noise injecting. are quite limited in terms of creating variation. As a result, due to the limitations of the currently available approaches, the proposed AnoMalNet architecture in this paper can be useful.

\section{Method}
Classifying malaria cell images into either parasite infected cells or uninfected cell is a binary classification task. Traditional deep neural network (DNN) models can be used for solving this problem. In addition to this, it can be easily formulated as an outlier detection problem with the help of auto-encoders. In the following subsections, several DNN models like LeNet, VGG16, ResNet50, MobileNetv2, and autoencoders are described.

\subsection{Deep neural network models}
One of the earliest DNN models that have been proposed is LeNet \cite{b23}. It consists of three convolutional layers of kernel size 5 and two average pooling layers. Additionally, it also has two fully-connected layers which act as the classifiers. Compared to LeNet, VGG16 is a much deeper model comprising 13 convolutional layers and 3 fully-connected layers \cite{b24}. This model uses a smaller kernel size. In this approach, the kernel size is set to 3. Theoretically, a deeper neural network model should perform better than a shallow one. However, it was found that if a deep neural network is used then a problem arises which is the vanishing gradient problem. In order to solve this problem, the ResNet model was proposed which contains residual blocks \cite{b25}. These blocks contain skip connections that take the output from previous layers and feed it to the later ones. Thus it helps in solving the vanishing gradient problem. MobileNetV2 is a CNN architecture that is specifically designed for mobile devices. It is built on an inverted residual structure where the bottleneck layers are coupled by residual connections \cite{b26}. Lightweight depthwise convolutions are used in the intermediate expansion layer as a source of non-linearity to filter features.

An autoencoder is a DNN model which tries to learn its input to the output. It is a two part DNN model where the first part is an encoder network while the later one is a decoder network. The task of the encoder part is to encode the input to a representation of the input of a smaller dimension while the decoder tries to collect this representation and reconstruct it to the original input. With the help of this encoder and decoder network, an autoencoder can be used to create a spare representation of the input data. For this research work a custom convolutional autoencoder is used. The major difference between this AnoMalNet architecture and autoencoders is that, in convolutional autoencoders, convolutional layers are used while in the later case, regular feed forward neural networks are used. In order to create the encoder network three convolution layers were used of 4, 16, and 32 channels respectively. The kernel size for all of these convolution layers were set to 3$\times$3 and the padding was set to 1. All of these convolution layers were followed by relu activation function and max pooling layers of 2$\times$2. 

\begin{figure}[H]
    \centering
    \includegraphics[scale=0.45]{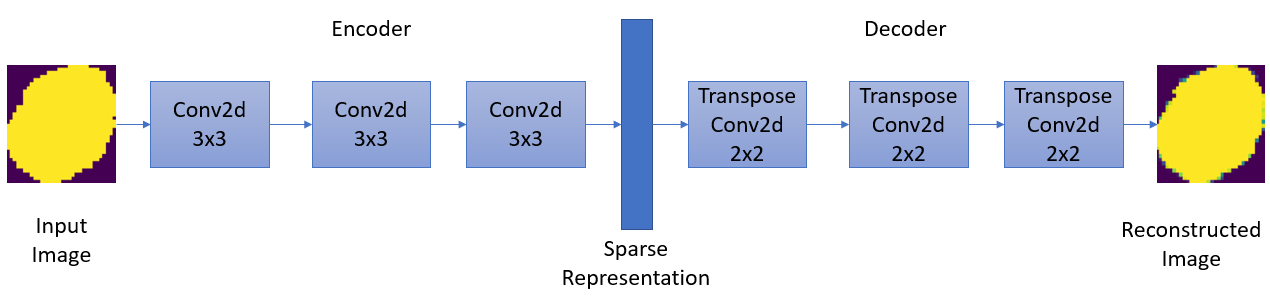}
		\vspace{.7em}
    \caption{Custom autoencoder}
    \label{fig:model}
\end{figure}

The decoder network was comprised of three transpose convolution layers of 32, 16, and 4 channels respectively. In this case, 2$\times$2 was set as the kernel size and the stride value was set at 2. Apart from the last transpose convolution layer, all the other layers' outputs were passed through a relu function while the output of the final layer went through a sigmoid activation function. Figure \ref{fig:model} provides a graphical view of our custom autoencoder. 

\subsection{Proposed approach} \label{sub:proposed_approach}
At the very beginning, some simple data preprocessing techniques are applied on the image dataset. All the images are reshaped to a dimension of 32$\times$32. Additionally, all the images are converted from the red green blue (RGB) to gray-scale color space. Changing the color space domain doesn't create many problems for this task because the parasite is still visible in the gray-scale color space and by doing so it removes unnecessary noises in the images.

\begin{figure}[H]
\centering
  \subfloat[]{
	\begin{minipage}[c][]{0.5\textwidth}
	   \centering
	   \includegraphics[scale=0.45]{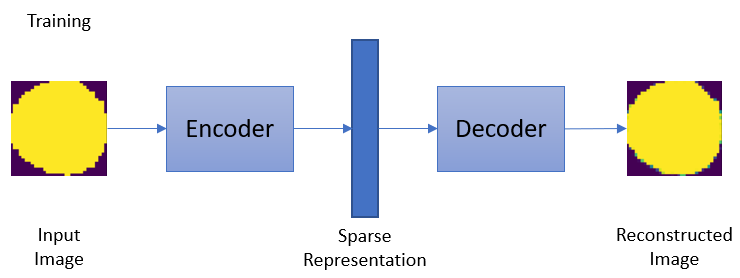}
	\end{minipage}}
 \hspace{-.8cm}		
  \subfloat[]{
	\begin{minipage}[c][]{0.5\textwidth}
	   \centering
	   \includegraphics[scale=0.45]{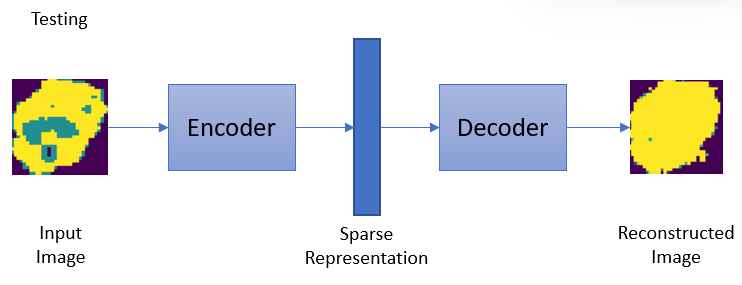}
	\end{minipage}}
	\vspace{.7em}
\caption{Proposed methodology (a) uninfected cell images for training and (b) infected cell images used for testing only}
    \label{fig:methodology}
\end{figure}

After this step, a custom auto-encoder with the decoder network is trained using only uninfected cell images. Mean squared error (MSE) is used as a loss function to train the weights of the model. The proposed approach is based on the intuition that, during testing, this trained auto-encoder will achieve a loss score for uninfected cell images. However, in the case of parasite infected cells, the model will output a significantly higher loss value. This can be visualized with the help of Figure \ref{fig:methodology}. The training and testing of the model in the case of uninfected cell images are shown while in Figures \ref{fig:methodology}(a) and (b), the performance of the model during inference on infected cell images is shown. With the help of simple statistics, a cut-off point can be established to label unknown cell images as infected ones or uninfected ones. An unknown image is determined as an outlier or infected cell if the loss value of the unknown image which we get after passing through the model is more than the mean plus three times of the standard deviation of train loss. This is a standard statistical approach to determine whether a particular data is an outlier or not.

\begin{figure}[H]
\centering
  \subfloat[]{
	\begin{minipage}[c][]{0.5\textwidth}
	   \centering
	   \includegraphics[scale=0.6]{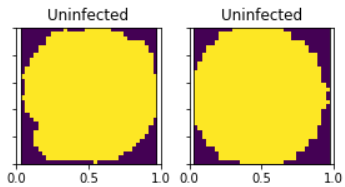}
	\end{minipage}}
 \hspace{-.8cm}		
  \subfloat[]{
	\begin{minipage}[c][]{0.5\textwidth}
	   \centering
	   \includegraphics[scale=0.6]{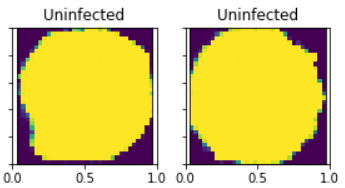}
	\end{minipage}}
	\hspace{-.8cm}
	\subfloat[]{
	\begin{minipage}[c][]{0.5\textwidth}
	   \centering
	   \includegraphics[scale=0.6]{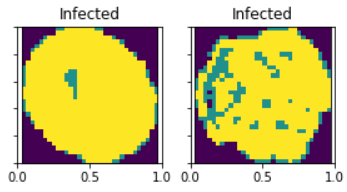}
	\end{minipage}}
 \hspace{-.8cm}		
  \subfloat[]{
	\begin{minipage}[c][]{0.5\textwidth}
	   \centering
	   \includegraphics[scale=0.6]{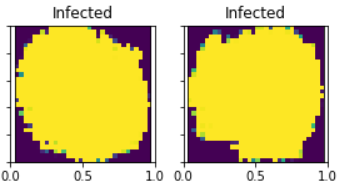}
	\end{minipage}}
	\vspace{.7em}
\caption{Visualization of (a) original uninfected cell, (b) reconstructed uninfected cell, (c) original infected cell, and (d) reconstructed infected cell} \label{fig:methodology_visualization}
\end{figure}

\newpage
Reconstruction of different types of cell images are shown in Figure \ref{fig:methodology_visualization}. Original and reconstructed images of uninfected cells are shown in Figures \ref{fig:methodology_visualization}(a) and (b). Additionally, Figures \ref{fig:methodology_visualization}(c) and (d) contain the original and reconstructed images of infected cells. From these figures, it can be seen that in the case of uninfected normal cell images, the reconstructed image is quite similar to the original one. However, in case of parasite infected cells, the reconstruction is not so good. The model is able to get the shape correct to some extent but not the parasite inside the cell. 

\section{Experimental Results and Analysis}
The models and proposed approach that were described in the previous section were used for experimentation on a dataset which can be used for malaria parasite classification. Details about the dataset can be found in the following subsection. Apart from this, a detailed discussion on the experimental setup, results and analysis are provided in the following subsections. 

\subsection{Dataset description}
The dataset that has been used here is collected from the National Institute of Health (NIH). It contains a total of 27,558 images. Among these, 13,779 images are infected with malaria parasites while the rest of them are uninfected cell images. All the images are in RGB color-space. 

\subsection{Experimental setup}
In order to train the AnoMalNet model, randomly selected 1,607 uninfected cell images were used. For validating the models performance, 407 uninfected cell images were used. A total of 4,009 images were used to train and evaluate the performance of the AnoMalNet model. The MSE loss function was used along with the Adam optimizer and the learning rate was set to 0.01. The model was trained for a total of 200 epochs. During the testing phase, a total of 5,512 images were used. Among them, 2,757 were parasite infected images and 2,755 were uninfected cells. After training the model, the loss value of the validation set was used to calculate the mean and standard deviation which was later used for creating a threshold for decision making. As mentioned in section 3.2, a particular image is labeled as a parasitized cell image if it's loss value is greater than the mean plus three times of the standard deviation. Using this threshold, all the images are classified with the help of an autoencoder.

Several DNN networks namely LeNet, VGG16, ResNet50, and MobileNetV2 were trained to compare the performance of the proposed approach \cite{b23}-\cite{b26}. All of these models were trained on 22,046 images from the dataset which contain both infected and uninfected cell images. These DNN models were trained for 100 epochs each using Adam optimizer while keeping the learning rate to 0.01, having cross entropy as the loss function.

\subsection{Results and discussion}
\begin{figure}[H]
\centering
  \subfloat[]{
	\begin{minipage}[c][]{0.5\textwidth}
	   \centering
	   \includegraphics[scale=0.45]{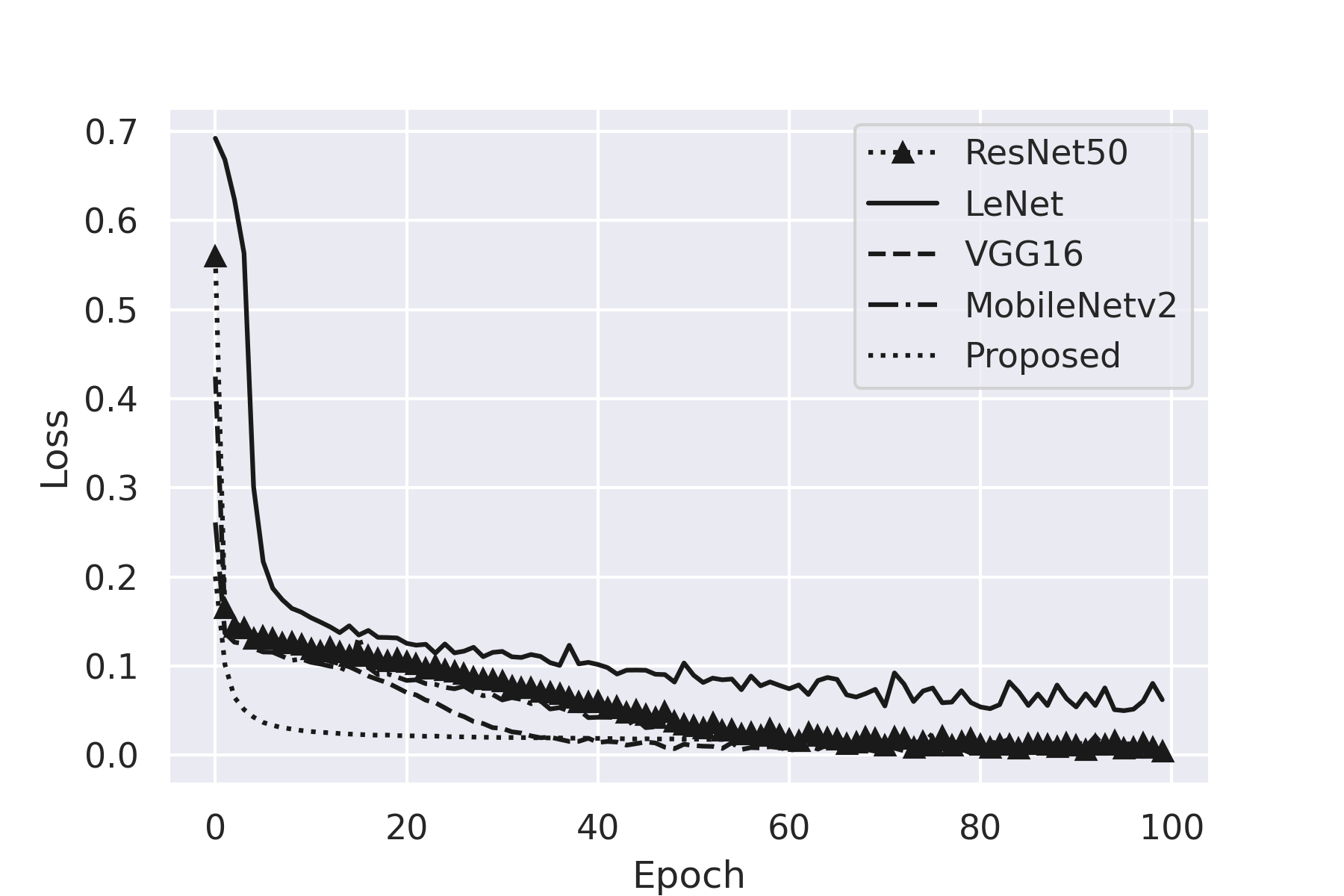}
	\end{minipage}}
 \hspace{-.8cm}		
  \subfloat[]{
	\begin{minipage}[c][]{0.5\textwidth}
	   \centering
	   \includegraphics[scale=0.45]{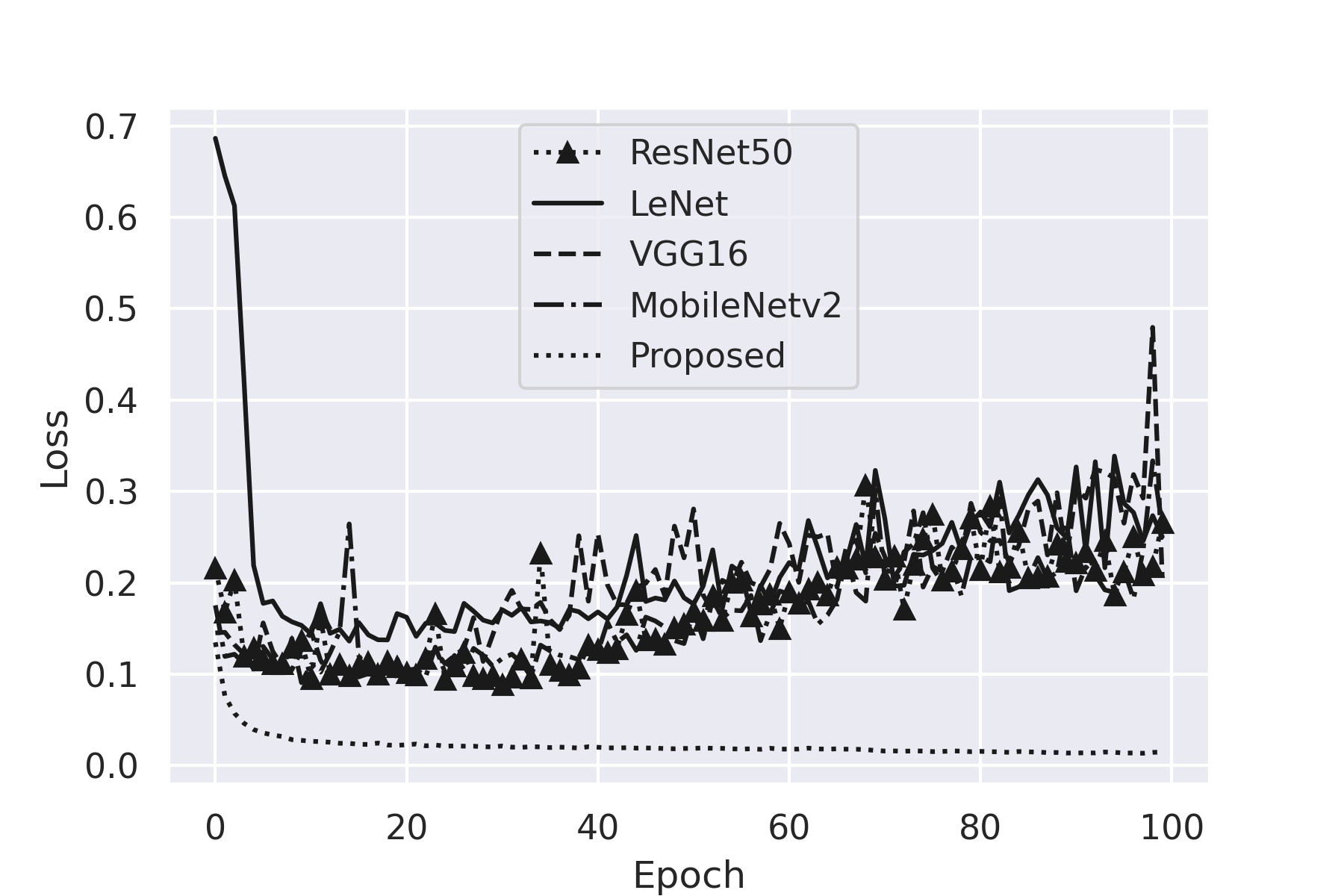}
	\end{minipage}}
	\vspace{.7em}
\caption{Loss vs epoch graph for (a) training data and (b) testing data}\label{fig:loss curves}
\end{figure}

For all the trained models, loss vs epoch curve can be found in Figure \ref{fig:loss curves}. In Figure \ref{fig:loss curves}(a) and (b) training and testing loss of all the models can be visualized. From this curve, it can be seen that after a while traditional DNNs tend move in such a direction that the test loss increases. However, autoencoder based proposed approach doesn't have this problem and moves toward achieving a zero score in test loss. Although it should be kept in mind that the autoencoder is testing on unknown uninfected cell images while the DNNs are performing tests on unknown infected and uninfected cell images.

\begin{figure}[H]
\centering
\includegraphics[scale=0.55]{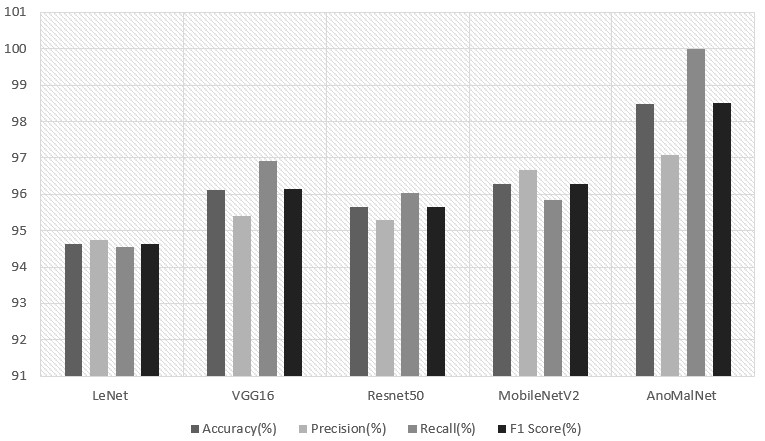}
\vspace{.7em}
\caption{A bar chart illustrating different models performance}
    \label{fig:bar}
\end{figure}

After training the model properly, the results shown in Table \ref{table:1} are obtained. In order to better understand the performance of the proposed approach four different metrics are used which are accuracy, precision, recall and F1 score. To be able to visualize the comparison of various models a bar chart is displayed as well in Figure \ref{fig:bar}. The lowest performing model according to the bar chart and the table is LeNet. This model was able to acquire 94.64\% accuracy, 94.73\% precision, 94.56\% recall and 94.64\% F1 score. MobileNetv2 attained the second highest accuracy and F1 score which is 96.28\% and 96.27\%. However, the best performing method was the proposed AnoMalNet. It was able to achieve 98.49\% accuracy, 97.07\% precision, 100\% recall and 98.52\%  F1 score.

\vspace{-1.7em}
\vspace{-.75em}
\begin{table}[H]
\caption{Comparative study of the performance of the proposed method against other DNN models}
\label{table:1}
\centering
\begin{tabular}{lcccc}
\hline
\multicolumn{1}{c}{Method} & Accuracy (\%)       & Precision (\%)      & Recall (\%)       & F1 score (\%)       \\ \hline
LeNet        & 94.64          & 94.73          & 94.56        & 94.64          \\ 
VGG16        & 96.11          & 95.40           & 96.91        & 96.14          \\ 
Resnet50    & 95.64          & 95.28          & 96.04        & 95.66          \\ 
MobileNetV2 & 96.28          & 96.67          & 95.84        & 96.27          \\
 
AnoMalNet     & \textbf{98.49} & \textbf{97.07} & \textbf{100} & \textbf{98.52} \\ \hline
\end{tabular}
\end{table}

\vspace{-.75em}

Apart from comparing with traditional DNN models, another study is also conducted with other proposed methods which are proposed by other researchers. Table  \ref{tab:2} shows a comparison of this manuscripts proposal with other research works. From this table, it can be seen that the proposed autoencoder based outlier detection method outperforms other traditional DNN based classification techniques.

\begin{table}[H]
\centering
\caption{Comparative study of the performance of the proposed method against other published approaches}
\label{tab:2}
\begin{tabular}{lc}
\hline
\multicolumn{1}{c}{Method} & Accuracy (\%) \\ \hline
Narayanan \emph{et al}. \cite{b13}     & 96.60    \\
Reddy and Juliet \cite{b14} & 95.40    \\ 
Raihan and Nahid \cite{b11}   & 94.78    \\
AnoMalNet                 & \textbf{98.49}    \\ \hline
\end{tabular}
\end{table}

\section{Conclusion}
An autoencoder-based DNN architecture is presented in this research work for classifying malaria parasite infections in cell images. This DNN model is trained to identify outliers i.e. parasite-infected cells. As this model is trained on completely normal cell images, this method provides an advantage in scenarios where disease-positive samples are scarce. With the help of MSE loss and a threshold, this approach can correctly identify images with malaria parasites. Additional comparisons with other traditional DNN models have been shown in this experiment from which it can be seen that the proposed approach performs better than traditional DNN models. There are some scopes of improvement in this research work. Like incorporating more complex datasets for this model, expanding the task from binary classification to multi-class classification.

\bibliographystyle{IEEEtran}

\begin{thebibliography} {99} 
\footnotesize
\itemsep 0pt 


\bibitem{b1}
WH0, “World malaria report 2022,” 2023. [Online]. Available: https://www.who.int/publications/i/item/9789240064898 (accessed Jan. 02, 2023).

\bibitem{b2}
M. K. Gourisaria, S. Das, R. Sharma, S. S. Rautaray, and M. Pandey, “A deep learning model for malaria disease detection and analysis using deep convolutional neural networks,” \emph{International Journal on Emerging Technologies}, vol. 11, no. 2, pp. 699–704, 2020. 


\bibitem{b3}
X. Liu, L. Song, S. Liu, and Y. Zhang, “A review of deep-learning-based medical image segmentation methods,” \emph{Sustainability}, vol. 13, no. 3, p. 1224, Jan. 2021, doi: 10.3390/su13031224.

\bibitem{b4}
H. Guan and M. Liu, “Domain adaptation for medical image analysis: a survey,” \emph{IEEE Transactions on Biomedical Engineering}, vol. 69, no. 3, pp. 1173–1185, Mar. 2022, doi: 10.1109/TBME.2021.3117407.

\bibitem{b5}
Q. Quan, J. Wang, and L. Liu, “An effective convolutional neural network for classifying red blood cells in malaria diseases,” \emph{Interdisciplinary Sciences: Computational Life Sciences}, vol. 12, no. 2, pp. 217–225, Jun. 2020, doi: 10.1007/s12539-020-00367-7.

\bibitem{b6}
A. Rogers, M. Gardner, and I. Augenstein, “QA dataset explosion: a taxonomy of nlp resources for question answering and reading comprehension,” \emph{ACM Computing Surveys}, vol. 55, no. 10, pp. 1–45, Oct. 2023, doi: 10.1145/3560260.

\bibitem{b7}
M. Birjali, M. Kasri, and A. Beni-Hssane, “A comprehensive survey on sentiment analysis: Approaches, challenges and trends,” \emph{Knowledge-Based Systems}, vol. 226, p. 107134, Aug. 2021, doi: 10.1016/j.knosys.2021.107134.

\bibitem{b8}
L. Schoneveld, A. Othmani, and H. Abdelkawy, “Leveraging recent advances in deep learning for audio-visual emotion recognition,” \emph{Pattern Recognition Letters}, vol. 146, pp. 1–7, Jun. 2021, doi: 10.1016/j.patrec.2021.03.007.

\bibitem{b9}
D. Michelsanti \emph{et al}., “An overview of deep-learning-based audio-visual speech enhancement and separation,” \emph{IEEE/ACM Transactions on Audio, Speech, and Language Processing}, vol. 29, pp. 1368–1396, 2021, doi: 10.1109/TASLP.2021.3066303.

\bibitem{b10}
W. Lee, J. J. Seong, B. Ozlu, B. S. Shim, A. Marakhimov, and S. Lee, “Biosignal sensors and deep learning-based speech recognition: a review,” \emph{Sensors}, vol. 21, no. 4, p. 1399, Feb. 2021, doi: 10.3390/s21041399.

\bibitem{b11}
M. J. Raihan and A.-A. Nahid, “Malaria cell image classification by explainable artificial intelligence,” \emph{Health and Technology}, vol. 12, no. 1, pp. 47–58, Jan. 2022, doi: 10.1007/s12553-021-00620-z.

\bibitem{b12}
S. M. Lundberg and S.-I. Lee, “A unified approach to interpreting model predictions,” in \emph{NIPS’17: Proceedings of the 31st International Conference on Neural Information Processing Systems}, 2017, pp. 4768–4777.

\bibitem{b13}
B. N. Narayanan, R. A. Ali, and R. C. Hardie, “Performance analysis of machine learning and deep learning architectures for malaria detection on cell images,” in \emph{Applications of Machine Learning, Sep. 2019, p. 29, doi: 10.1117/12.2524681.
}
\bibitem{b14}
A. S. B. Reddy and D. S. Juliet, “Transfer learning with ResNet-50 for malaria cell-image classification,” in \emph{2019 International Conference on Communication and Signal Processing (ICCSP)}, Apr. 2019, pp. 0945–0949, doi: 10.1109/ICCSP.2019.8697909.

\bibitem{b15}
S. Rajaraman \emph{et al}., “Pre-trained convolutional neural networks as feature extractors toward improved malaria parasite detection in thin blood smear images,” \emph{PeerJ}, vol. 6, p. e4568, Apr. 2018, doi: 10.7717/peerj.4568.

\bibitem{b16}
J. A. Quinn, R. Nakasi, P. K. B. Mugagga, P. Byanyima, W. Lubega, and A. Andama, “Deep convolutional neural networks for microscopy-based point of care diagnostics,” in \emph{Proceedings of the 1st Machine Learning for Healthcare Conference}, 2016, pp. 271–281

\bibitem{b17}
D. Bibin, M. S. Nair, and P. Punitha, “Malaria parasite detection from peripheral blood smear images using deep belief networks,” \emph{IEEE Access}, vol. 5, pp. 9099–9108, 2017, doi: 10.1109/ACCESS.2017.2705642.

\bibitem{b18}
S. Lipsa and R. K. Dash, “MalNet – an optimized CNN based method for malaria diagnosis,” in \emph{2022 2nd International Conference on Intelligent Technologies (CONIT)}, Jun. 2022, pp. 1–6, doi: 10.1109/CONIT55038.2022.9848328.


\bibitem{b19}
H. A. Nugroho and R. Nurfauzi, “GGB color normalization and faster-RCNN techniques for malaria parasite detection,” in \emph{2021 IEEE International Biomedical Instrumentation and Technology Conference (IBITeC)}, Oct. 2021, pp. 109–113, doi: 10.1109/IBITeC53045.2021.9649152.


\bibitem{b20}
C. K. Tan, C. M. Goh, S. A. Z. B. S. Aluwee, S. W. Khor, and M. T. Chai, “Malaria parasite detection using residual attention U-net,” in \emph{2021 IEEE International Conference on Signal and Image Processing Applications (ICSIPA),} Sep. 2021, pp. 122–127, doi: 10.1109/ICSIPA52582.2021.9576814.

\bibitem{b21}
G. Mariani, F. Scheidegger, R. Istrate, C. Bekas, and C. Malossi, “BAGAN: data augmentation with balancing GAN,” arXiv:1803.09655, Mar. 2018, [Online]. Available: http://arxiv.org/abs/1803.09655.

\bibitem{b22} 
C. Shorten and T. M. Khoshgoftaar, “A survey on image data augmentation for deep learning,” \emph{Journal of Big Data}, vol. 6, no. 1, p. 60, Dec. 2019, doi: 10.1186/s40537-019-0197-0.

\bibitem{b23} 
Y. LeCun, “LeNet-5, convolutional neural networks.” http://yann.lecun.com/exdb/lenet (accessed Jan. 02, 2023).

\bibitem{b24}
K. Simonyan and A. Zisserman, “Very deep convolutional networks for large-scale image recognition,” in \emph{3rd International Conference on Learning Representations (ICLR 2015)}, 2015, pp. 1–14.


\bibitem{b25}
K. He, X. Zhang, S. Ren, and J. Sun, “Deep residual learning for image recognition,” in \emph{2016 IEEE Conference on Computer Vision and Pattern Recognition (CVPR)}, Jun. 2016, pp. 770–778, doi: 10.1109/CVPR.2016.90.


\bibitem{b26}
M. Sandler, A. Howard, M. Zhu, A. Zhmoginov, and L.-C. Chen, “MobileNetV2: inverted residuals and linear bottlenecks,” in \emph{2018 IEEE/CVF Conference on Computer Vision and Pattern Recognition}, Jun. 2018, pp. 4510–4520, doi: 10.1109/CVPR.2018.00474.

 \end{thebibliography}

\section*{BIOGRAPHIES OF AUTHORS}
\small
\begin{biography}[{\includegraphics[width=2.5cm,height=3cm,clip,keepaspectratio]{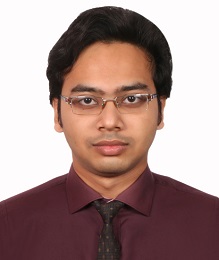}}]
\textbf{Aminul Huq} 
\href{https://orcid.org/0000-0002-6500-6097}{\includegraphics[width=0.02\textwidth]{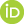}} 
\href{https://scholar.google.com/citations?user=JTl6Bs8AAAAJ&hl=en}{\includegraphics[width=0.02\textwidth]{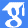}}
\href{}{\includegraphics[width=0.02\textwidth]{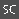}}
\href{}{\includegraphics[width=0.02\textwidth]{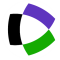}}
is a lecturer at Brac University. He received his master's in computer science and technology from Tsinghua University in 2021. He completed his bachelor's from Rajshahi University of Engineering and Technology in 2017. His research interest lies in the field of multi-task learning and deep learning. He can be contacted at email: aminul.huq@bracu.ac.bd. 
\vspace{.7em}
\end{biography}

\begin{biography}[{\includegraphics[width=2.5cm,height=3cm,clip,keepaspectratio]{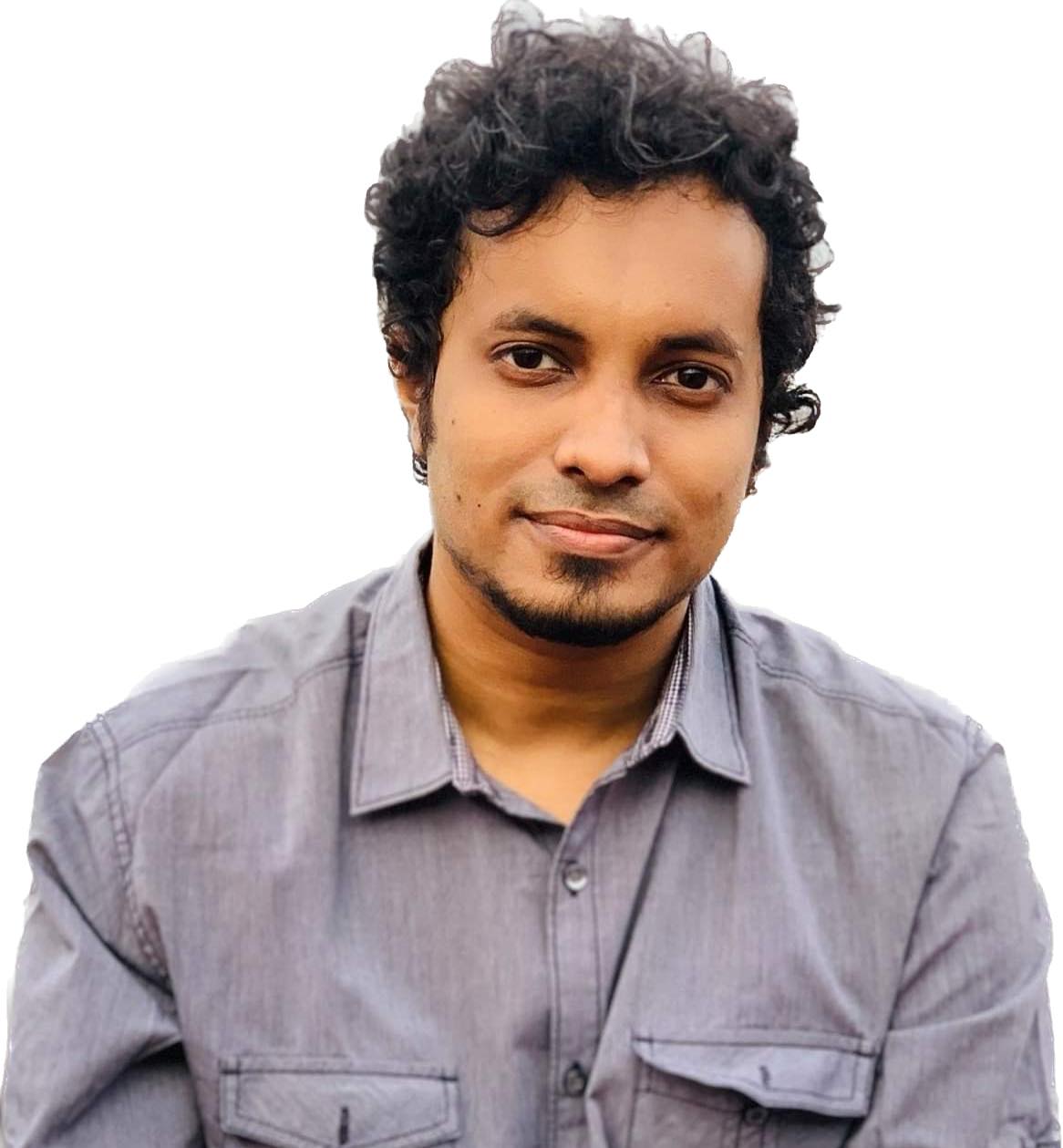}}]
\textbf{Md Tanzim Reza} 
\href{https://orcid.org/0000-0001-8964-1565}{\includegraphics[width=0.02\textwidth]{orcid.png}} 
\href{https://scholar.google.com/citations?user=hnmuI_IAAAAJ&hl=en&oi=ao}{\includegraphics[width=0.02\textwidth]{gscholar.png}}
\href{}{\includegraphics[width=0.02\textwidth]{scopus.png}}
\href{}{\includegraphics[width=0.02\textwidth]{Clarivate.png}}
completed his B.Sc. in computer science and engineering from BRAC University in 2018. Afterward, he joined as a contractual lecturer in 2019 and eventually became full-time lecturer in January 2020. His research interest is artificial intelligence, machine learning, natural language processing, and computer vision. He can be contacted at email: rezatanzim@gmail.com.
\end{biography}

\begin{biography}[{\includegraphics[width=2.5cm,height= 3 cm,clip,keepaspectratio]{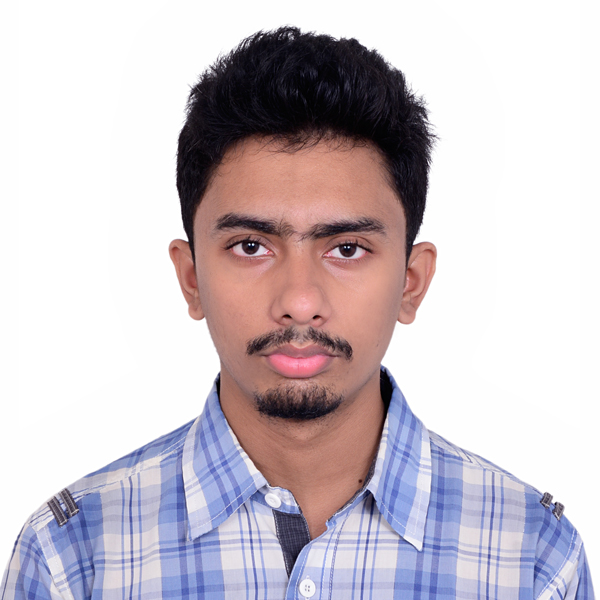}}]
\textbf{Shahriar Hossain } 
\href{https://orcid.org/0000-0002-6132-1305}{\includegraphics[width=0.02\textwidth]{orcid.png}} 
\href{https://scholar.google.com/citations?user=UL1s_poAAAAJ&hl=en}{\includegraphics[width=0.02\textwidth]{gscholar.png}}
\href{}{\includegraphics[width=0.02\textwidth]{scopus.png}}
\href{}{\includegraphics[width=0.02\textwidth]{Clarivate.png}}
earned his B.Sc. in computer science and engineering from BRAC University in 2021. He is currently working as a research assistant at the Department of Computer Science and Engineering at BRAC University. He has attended several national and international competitions in the field of robotics. His current research interests include machine learning, deep learning, and robotics. He can be contacted at email: shahriar.hossain@bracu.ac.bd.
\end{biography}

\begin{biography}[{\includegraphics[width=2.5cm,height=4cm,clip,keepaspectratio]{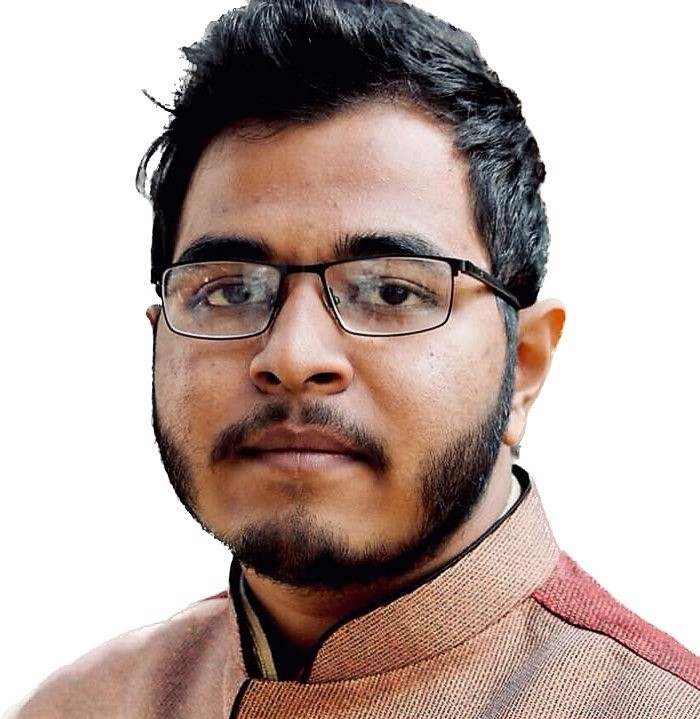}}]
\textbf{Shakib Mahmud Dipto} 
\href{https://orcid.org/0000-0003-2704-118X}{\includegraphics[width=0.02\textwidth]{orcid.png}} 
\href{https://scholar.google.com/citations?user=tHqA-0QAAAAJ&hl=en}{\includegraphics[width=0.02\textwidth]{gscholar.png}}
\href{}{\includegraphics[width=0.02\textwidth]{scopus.png}}
\href{}{\includegraphics[width=0.02\textwidth]{Clarivate.png}}
is working as a research assistant at BRAC University. He received his bachelor of science degree in computer science and engineering from BRAC University in 2021. His research interests lie in the fields of computer vision, medical image processing, machine learning, and deep learning. He can be contacted at email: diptomahmud2@gmail.com.
\end{biography}

\end{document}